# Porous aluminium decorated with rhodium nanoparticles – preparation and use as platform for UV plasmonics

Shrobona Banerjee[a,b], Luca Mattarozzi[c], Nicolò Maccaferri[d], Sandro Cattarin[c], Shukun Weng[a], Ali Douaki[a], German Lanzavecchia[a], Anastasiia Sapunova[a], Francesco D'Amico[e], Qifei Ma[a,g], Yanqui Zou[a,h], Roman Krahne[a], Janina Kneipp[b] and Denis Garoli[a,f,g*]

There is a high current interest for novel plasmonic platforms and materials able to extend their applicability into the ultraviolet (UV) region of the electromagnetic spectrum. In the UV it is possible to explore spectral properties of biomolecules with small cross section in the visible spectral range. However, most used metals in plasmonics have their resonances at wavelengths > 350 nm. Aluminum and rhodium are two interesting candidate materials for UV plasmonics, and in this work we developed a simple and low-cost preparation of functional substrates based on nanoporous aluminum decorated with rhodium nanoparticles. We demonstrate that these functionalized nanoporous metal films can be exploited as plasmonic materials suitable for enhanced UV Raman spectroscopy.

## Introduction

Nucleic acid-based aptamers are single-stranded, short DNA or RNA sequences, isolated *in vitro* from a library of synthetic Plasmonics born in the 1950s with the discovery of grating anomalies by Wood. After this, the excitation of surface plasmon polaritons (SPPs) by randomly rough surfaces has initiated fascinating topics of modern physics like enhanced backscattering or Anderson localization of photons. Then, plasmonics went through a novel impulsion in mid-1970s when the surface-enhanced Raman scattering (SERS) was discovered.[1–3] Ultimately, in the last two decades a a new boom of plasmonics and its applications has occurred. Nowadays surface plasmons find extensive applications in SERS, metal-enhanced fluorescence (MEF), photocatalysis, optical trapping, thermal effects, etc.[2] Thanks to the innovation in fabrication techniques and spectroscopic methods, plasmonics can now find applications in multiple spectral regions. The most used noble metals (Au and Ag) are typically applied in the visible/near-infrared (Vis/NIR), while metals such as aluminum, magnesium, rhodium and gallium can be used to extend their use to higher energy (into the UV and deep-UV wavelengths)[4,5]. The excitation of biomolecules using UV radiation is in particular interesting for Raman spectroscopy. In fact, some biomolecules have small Raman cross sections in the visible and NIR regions[6,7] and the use of higher energy to excite them can increase the detection limit [8–12]. Here aluminum (Al) is the most extensive explored material for UV plasmonics [13]. In order to use Al to generate localized surface plasmon resonances in the UV, very small metallic features/nanostructures must be prepared, for example by means of advanced lithographic techniques or chemical synthesis of nanoparticles [14,15]. As an alternative approach to produce nanostructured metals, the preparation of Al films as nanoporous material (NPM) has been recently reported[7,16,17]. During the last decade, NPMs have attracted increasing interest due to their unique very high specific surface area, and their applications in multiple fields is extensively documented[18].Porous Al structures (as metal with low oxygen contents) can be prepared by means of galvanic replacement reaction or electrochemical dealloying, respectively, starting from an alloy of $Mg_3Al_2$.[7,19] An alternative method to obtain micro and nanoporous Al films is based on the electroless treatment known as the Galvanic Displacement (GD) reaction. GD may be used with simple chemical apparatus to deposit metallic dots or thin layers spontaneously at open circuit voltage. The approach is applicable with apparent advantages to: i) metallic or semiconducting substrates of complex geometry, as the solution may creep into recessed but connected empty volumes; ii) topologies like patterned planar substrates, as the procedure may skip the application of masks. Due to these advantages, it is the object of intense current investigation[20].

In our case, a metal of low redox potential E like Al, oxidizes and dissolves

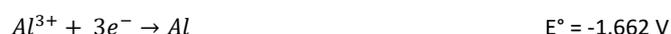

$$Al^{3+} + 3e^- \rightarrow Al \qquad\qquad E° = -1.662\ V$$

whereas the ions of a comparatively more noble metal ($Rh^{3+}$) reduce and deposit

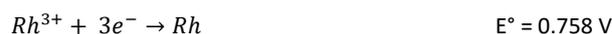

$$Rh^{3+} + 3e^- \rightarrow Rh \qquad\qquad E° = 0.758\ V$$

The presence in solutions of complexing ions like $Cl^-$ may alter the individual potentials but does not change the relative positions in the redox scale. The overall process

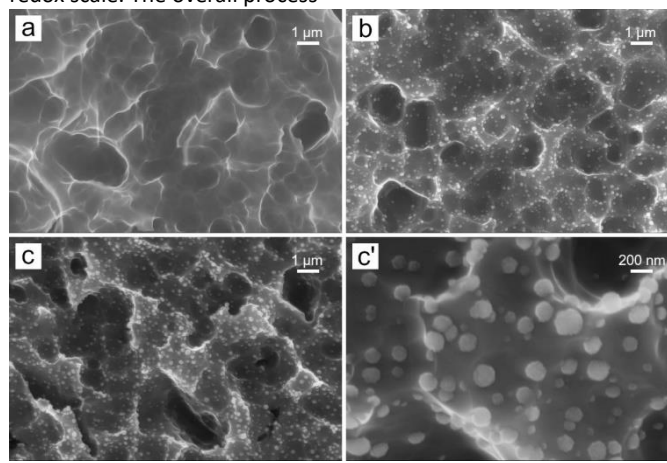

$$Al + Rh^{3+} \rightarrow Al^{3+} + Rh$$

may cause formation of Rh deposits of various dimensions depending on treatment duration and operating conditions.

As previously mentioned, rhodium (Rh) is another interesting candidate metal for UV plasmonics and examples of its use for enhanced spectroscopies have been reported.[8,21–23] The current main limitation in the use of Rh in the preparation of nanostructures is its very high cost. Therefore, alternative low costs approaches are sought. Also, the combination of two plasmonic metals in the same platform can, in principle, provide more tunable optical properties and better field enhancement for spectroscopic applications.[24,25]

In this paper, we discuss the facile and low cost preparation of porous Al films decorated with Rh nanoparticles with increasing surface coverage density. In view of potential application as UV plasmonic platform, we tested the films for SERS using an excitation wavelength of 266 nm to probe adenine molecules uniformly deposited on the surface. The results are discussed with the support of numerical simulations to rationalize the observed trend in Raman signal intensities.

## Results and Discussion

The morphology of porous Al samples decorated with Rh nuclei on the surface, obtained by galvanic displacement process, was characterized by SEM. Merely etched samples (Figure 1a) show a strongly roughened/porous Al surface with features of few to several microns size, presumably reflecting a metallurgic grain structure. Upon GD treatment, randomly distributed Rh nanoparticles appear (Fig. 1b-c-c'), with a morphology approaching a hemispherical cap. Moreover, for a given $Rh^{3+}$ concentration, with increasing time of treatment $t_{GD}$, a rise in the number of particles and their average dimensions was observed, reaching typical diameters in the range 100-300 nm for $t_{GD}$=2 min. On the other hand, exposing for the same time $t_{GD}$=1 min Al surfaces to different $Rh^{3+}$ concentrations (samples 05-1 and 20-1 in Table 1), the sample 20-1 exposed to the larger [$Rh^{3+}$] shows a much larger density of nucleated particles, and a limited increase in their average dimensions, as the particles tend to approach overlapping and complete substrate coverage.

Values of Rh at% reported in Table 1 were obtained as panoramic averaged values, at x1000 of magnification, and do not provide a reliable quantitative analysis due to the inhomogeneous depth profile of the samples. However, samples from 05-1 to 05-6 show a significant qualitative trend of increasing Rh content with increasing time of GD treatment. Data from the last sample 20-1 shows that the bath containing 2.0 mM $Na_3RhCl_6$ concentration causes a rather fast deposition, somewhat difficult to control and standardize.

Building on previous studies that have demonstrated the potential of nanoporous aluminum (NPA) to function as plasmonic substrates for SERS[19], the rhodium nanoparticles-coated NPA described here was used in Raman experiments. Adenine was chosen as an analyte, as its strong absorption near the excitation wavelength can lead to additional (electronic) resonant enhancement in its Raman spectra[26–28]. In a previous work we successfully exploit adenine to test the UV-SERS enhancement effect of nanoporous Al substrates [7].

**Figure 1**. SEM images of a) untreated porous Al (Sample 00-0, no GD) and b-c-c') porous Al samples submitted to galvanic displacement in 0.5 mM Na3RhCl6, pH 2.0, for different times: b) tGD = 1 min (Sample 05-1); c) tGD = 2 min (Sample 05-2); c') tGD = 2 min, magnified detail.

Spectra obtained from all the samples are reported in Figure 2. All show clearly distinguishable bands corresponding to the molecular vibrations of adenine as reported previously in resonant Raman experiments with adenine and related compounds[29–31], as well as in pre-resonant SERS spectra on Rh composite materials[32,33].

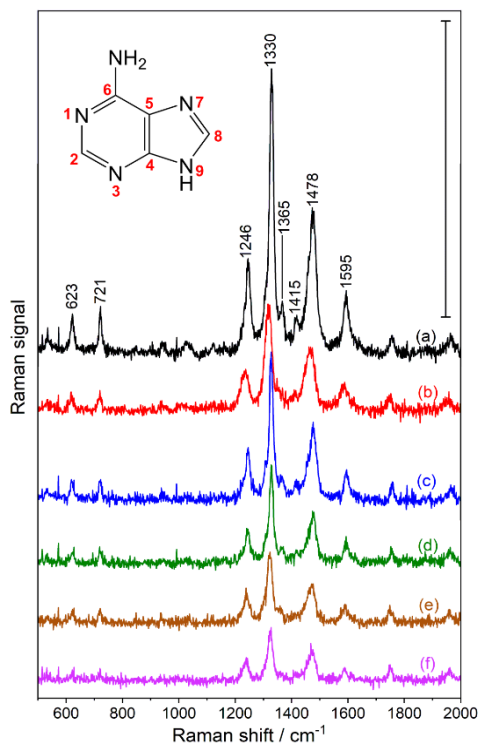

**Figure 2**. Averages of 6 spectra collected over 30 minutes for (a) untreated nanoporous Al (Sample 00-0); nanoporous Al subjected to galvanic displacement in 0.5 mM $Na_3RhCl_6$, pH 2.0 for (b) 1 min (Sample 05-1), (c) 2 min (Sample 05-2), (d) 1 min (Sample 05-4), (e) 6 min (Sample 05-6) and (f) in 2.0 mM $Na_3RhCl_6$, pH 2.0 for 1 min (Sample 20-1). Scale bar corresponds to 1 count per second (cps). Acquisition time per spectrum: 300 s, excitation wavelength: 266 nm, excitation intensity: 3.8 W cm$^{-2}$.

All observed bands are listed in Table 2, the inset in Figure 2 shows numbering of the atoms referred to in the assignment table. The skeletal vibrations corresponding to the six-membered pyrimidine ring (1592 cm$^{-1}$ and 1474 cm$^{-1}$)[30,34,35], and the five-membered imidazole (1330 cm$^{-1}$) indicate that the purine structure of the adenine molecules remains intact when deposited on the substrate in the thermal evaporation procedure. This band dominates the spectra, in agreement with previous work discussing a charge transfer-based chemical enhancement of this mode for adenine adsorbed on Rh under pre-resonant conditions[32]. The amino group is represented by the pronounced C6–$NH_2$ stretching vibration at 1246 cm$^{-1}$, also in agreement with previous works[26,30,34,35].

**Table 2**. Bands in the Raman spectra of adenine on the substrates and their tentative assignments, based on the listed references. The numbering of atoms in the adenine molecule is shown in the inset of Figure 2.

| Raman band (cm$^{-1}$) | Tentative assignment | Reference |
|---|---|---|
|   |   |   |

| 622 | Ring deformations C4-C5-N7, C4-C5-C6 | Ref. [35] |
| 721 | Purine ring breathing | Ref. [31] |
| 1246 | Stretching C6-$NH_2$, C-N7; Rocking $NH_2$ | Ref. [30,34] |
| 1330 | Stretching C5-N7, N1-C2; Bending C2-H, C8-H | Ref. [35] |
| 1365 | Stretching C8-N9, C4-N9; Bending C2-H, N9-H | Ref. [30,35] |
| 1415 | Stretching C4-N9, C8-N7, C6-N1 | Ref. [31], [35] |
| 1478 | Scissoring $NH_2$; $e_{1u}$ of six-membered ring (pyridine) | Ref. [35], [30,34] |
| 1595 | Scissoring $NH_2$; $e_{2g}$ of six-membered ring (pyridine) | Ref. [30,35] |

Considering the small amount of analyte in the few-nm layer that is deposited, the porous aluminum substrate could serve as efficient optical substrate that can carry small analyte amounts at its enlarged surface, and more importantly, could provide optimal plasmonic properties for potential SERS enhancement[32,36–40] as well. In order to investigate the influence of the presence of the rhodium nanostructures (Figure 1) on the properties as substrate for Raman experiments, the observed signal intensities can be compared. Interestingly, the overall signal intensity decreases for substrates containing higher proportions of RhNP (Figure 2b through 2f), and the control sample that does not contain any rhodium nanoparticles (Figure 2a) shows the highest signal. This result is counterintuitive, since one could expect an additional enhancement due to Rh NPs. We attribute the observed decreasing intensity to different effects. Firstly, while the NPA substrate itself may provide advantageous plasmonic properties for the SERS enhancement[19], it also provides a large available surface to the molecules. The presence of the Rh nanoparticles and their favorable possibilities of a chemical SERS enhancement[32] are 'outweighed' by the decreasing available surface on the aluminum substrate. More importantly, as discussed below, the regions of high electromagnetic field enhancement that are predicted to occur at the interface of the NPA and RhNPs (Fig. 4). Therefore, considering the deposition technique used to evaporate adenine (thermal evaporation is a directional process), the regions where the electromagnetic field is enhanced are inaccessible for the molecules. In order to take advantage of the plasmonic properties of the nanoparticles and the regions of high local optical fields (Fig. 4), we propose the deposition of the adenine or other analyte molecules prior to the RhNP layer on the NPA in future applications.

One of the main challenges to optical probing of organic compounds by excitation in the UV is their photo-degradation, which is caused by resonance with a multitude of electronic transitions. Figure 3 shows the signal at 1330 cm$^{-1}$ in 30 spectra of adenine collected consecutively over a period of 15 minutes. The mode at 1330 cm$^{-1}$ corresponds to a skeletal vibration of the pyrimidine ring[35] and its intensity could therefore be indicative of decomposition of the molecule. There is a decrease of the signal at 1330 cm$^{-1}$ of the adenine samples over time for all samples. The decrease in intensity of this band could indicate dissociation of skeletal purine ring bonds, and/or a re-orientation of the molecule on the nanostructure. For the sample without nanoparticles, we observed a rapid exponential decay in the signal intensity at 1330 cm$^{-1}$ (Figure 3A), followed by a slower decrease of the signal that is found for all samples. This is probably due to the structure of the unmodified NPA that results in optical properties that can accelerate the photo-damage of the deposited analyte. For the samples containing RhNPs, the decay occurs at lower rate (Figure 3B-F). From a linear fit of the signal intensity we see that the rate of decrease is reduced for samples with higher content of RhNPs (Figure 3B-F). We suggest that in this case the presence of the RhNPs, which may hinder access of a fraction of the molecules to the underlying NPA helps to prevent the rapid initial photo-decomposition. Therefore the architecture of the NPA containing RhNPs can be further optimized to ensure better metal-molecule interactions.

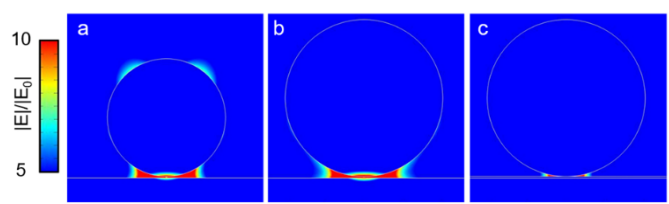

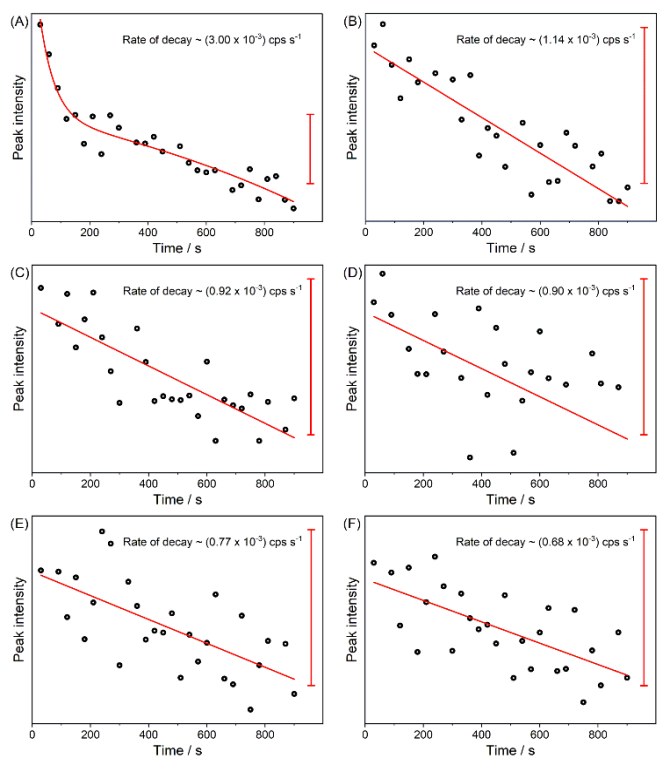

**Figure 3.** Signal intensities of the band at 1330 cm$^{-1}$ for (A) untreated nanoporous Al (Sample 00-0); nanoporous Al subjected to galvanic displacement in 0.5 mM $Na_3RhCl_6$, pH 2.0 for (B) 1 min (Sample 05-1), (C) 2 min (Sample 05-2), (D) 1 min (Sample 05-4), (E) 6 min (Sample 05-6) and (F) in 2.0 mM $Na_3RhCl_6$, pH 2.0 for 1 min (Sample 20-1). Scale bar corresponds to 1 count per second (cps). Acquisition time per spectrum: 300 s, excitation wavelength: 266 nm, excitation intensity: 0.8 W cm$^{-2}$.

To gain deeper insight in our experimental results, we performed finite elements model simulations using Comsol Multiphysics on a slightly simplified geometry for the Rh functionalized NPA system. We consider two experimental diameters / sizes of the obtained Rh NPs (150 and 200 nm) on top of a uniform Al layer.
Under illumination with an external source at 266 nm of the Rh NPs, the obtained electromagnetic field intensities (Fig. 5) resemble the well-known particle of a mirror configuration[41]. As can be observed, the field is mainly localized at the gap between the NPs and the metallic film below. In our experiments the adenine molecules were evaporated by means of sublimation directly on the surface of the Rh NPs / Al films. With such a directional deposition, it is reasonable to expect that the molecules were deposited on the upper surface of the NPs and on the exposed Al surface. Therefore the molecules are not present in the gap between the NPs and the Al metal, which explains the lower SERS signals that we observed.

**Figure 4.** FEM simulations of a Rh nanoparticle deposited on an Al film. Excitation wavelength 266 nm. Electromagnetic field intensity. (a) Rh nanoparticle with a diameter of 150 nm; (b) Rh nanoparticle with a diameter of 200 nm; (c) Rh nanoparticle with a diameter of 200 nm on a film of Al considering an additional 2 nm $Al_2O_3$ layer.

## Conclusions

In this paper we presented a new facile preparation procedure for plasmonic substrates in the UV. The fabrication relies on galvanic displacement reactions both for Al and Rh, two metals that are known for their interesting plasmonic properties in the UV spectral range. The plasmonic substrates consist of a porous Al film covered with Rh NPs, and the density of NP coverage can be easily controlled by adjusting the preparation parameters. Such combination of a porous Al film with Rh NPs is highly promising for applications in enhanced spectroscopies, in particular in UV-SERS.

## Experimental

**Sample preparation**

Square Al sheets 1 cm x 1 cm were cut from a commercial Al foil (99.99% Alfa Aesar, 1.5 mm thick). Polishing was performed using emery papers P1000 on both sides and then P4000 only on the side designated for the galvanic displacement (GD) process, until a homogeneous surface was achieved. At this stage the samples observed at SEM-EDS showed grooves due to mechanical scratch and inclusions of SiC particles. Al sheets were then submitted to ultrasonic treatment in water for 15 minutes to remove most residual SiC abrasive particles and dried. Prior to the GD process, Al sheets were etched in 5.0 M $H_2SO_4$ at 80 °C for 10 minutes, in order to eliminate residual SiC particles and obtain a more regular, renewed Al surface. Samples were then washed in water for 15 minutes with an ultrasonic bath and dried in a nitrogen flux (Fig. Ya). A polyester adhesive tape (3M 8992) was used to seal the back side, ensuring that GD processes occurred only on the front face.

All GD processes were performed in solutions prepared from deionized water (by Elga-Veolia Purelab Pulse System ρ >18 MΩ cm) and high-purity chemicals: NaCl (99.5% Merck), $Na_3RhCl_6 \cdot 12 H_2O$ (Alfa Aesar) and HCl (37% Merck). The solutions were aged for 48 h at 60 °C to ensure better reproducibility, due to the slow approach to equilibrium of the process of speciation of Rh chlorocomplexes[42,43].

In preliminary trials the effect of pH was investigated. The use of pH 3 or higher caused slow Rh deposition and rather scattered results; pH 1 produced rather fast reaction with visible development of $H_2$ gas bubbles and rapid Rh deposition, with some difficulty to control the extent of the latter. Optimal behaviour was observed at pH 2 and reported data refer to this pH.

The series of samples considered in the following is reported in Table 1, where the experimental conditions of the GD treatment are given. The solutions – adjusted to pH 2 with HCl - were maintained at 25 °C, and deaerated with $N_2$ for 20 minutes prior to sample immersion. Each sample is identified by a double code that recalls solutions' compositions and time of GD ($t_{GD}$). Most samples were prepared using a 0.5 mM $Na_3RhCl_6$ pH 2 solution, that offered best conditions to control and standardise deposition. After the GD treatment, the samples were washed gently by immersion in ultrapure water, dried in a box saturated with $N_2$, and stored under vacuum.

**Table 1.** Experimental conditions of samples preparation.

| Sample | GD bath | $t_{GD}$ / min | Rh at% (EDS) |
|---|---|---|---|
| 00-0 | No treatment | - | - |
| 05-1 | 0.5 mM $Na_3RhCl_6$ + 0.09 M NaCl | 1 min | 0.23-0.29 |
| 05-2 | 0.5 mM $Na_3RhCl_6$ + 0.09 M NaCl | 2 min | 0.43-0.54 |
| 05-4 | 0.5 mM $Na_3RhCl_6$ + 0.09 M NaCl | 4 min | 0.93-1.16 |
| 05-6 | 0.5 mM $Na_3RhCl_6$ + 0.09 M NaCl | 6 min | 1.24-1.32 |
| 20-1 | 2.0 mM $Na_3RhCl_6$ + 0.09 M NaCl | 1 min | 1.48-1.53 |

*Scanning Electron Microscopy (*SEM) images were obtained with a Zeiss SIGMA instrument, equipped with a field emission gun, operating in high vacuum conditions at an accelerating voltage of 20kV. EDS analyses were performed with an Oxford X-MAX apparatus.

*Energy-dispersive spectroscopy (EDS)* was performed within a JEOL JSM-7500LA SEM (JEOL, Tokyo, Japan), equipped with a cold field-emission gun (FEG), operating at 5 kV acceleration voltage. We measured the film composition through an Oxford instrument EDS setup (X-Max, 80 mm$^2$) Energy-dispersive spectroscopy (EDS, Oxford instrument, X-Max, 80 mm$^2$). The measurements were performed at 8 mm working distance, 5 kV acceleration voltage and 15 sweep count for each sample. EDS spectra from three different positions have been collected for each sample. In order to analyze the spectra, we used Aztec 1.2 software®, with automatic calibration of the standards and background subtraction. The instrument was calibrated with a Microanalysis Standard As-02756-AB 59 Metals & Minerals Carousel Serial HM, by SPI. For all the elements we analyzed the K-alpha lines. For each element, the composition percentages measured in different positions differed one from the other by less than 1%.

*Sample functionalization with adenine*

The substrates were deposited with a thin layer of adenine in a thermal evaporator by sublimating solid adenine under 4.0 x 10$^{-5}$ mbar. The film thickness was monitored in situ by a calibrated quartz crystal microbalance. The thickness of the final adenine layer is estimated to be about 2 nm.

*Raman spectroscopy:* Raman experiments with excitation in the ultraviolet (UV) were carried out at the Elettra Synchrotron radiation facility. A wavelength of 266 nm was used to probe the samples. The experimental setup consisted of a Czerny-Turner spectrograph of a collimator length of 750 mm, an entrance slit width of 100 μm, and a 1800 g/mm holographic grating that is coupled to a thermoelectrically cooled CCD detector. The spectral resolution was ~8 cm$^{-1}$ considering the full spectral range. Raman shift calibration was based on a spectrum of cyclohexane. From each sample, six spectra were collected consecutively with an acquisition time of 300 s per spectrum, yielding total respective collection times of 30 min per sample. The excitation intensity was adjusted to approximately 3.8 W cm$^{-2}$ for this set of experiments. Another set of data, comprising 30 spectra was collected at a lower intensity of ~0.8 W cm$^{-2}$ from each sample, yielding total collection times of 15 min per sample.

*FEM simulations* Numerical simulations were carried out to investigate the optical response of the Rh nanoparticles on the Al substrate. The electromagnetic response of an isolated Rh nanoparticle was simulated using the Finite-Element Method (FEM) implemented in the RF Module of Comsol Multiphysics®. The dimensions of the outer and inner diameters of the rings were set according to the average sizes obtained from the SEM images. The model computes electric field enhancement of the system. The unit cell was set to be 500 nm wide in both x- and y- and 1000 nm long in the z-direction, with perfect matching layers (200 nm thick) at the borders.

## Author Contributions

SB, SW, QM, FDA, JK and YZ performed the Raman measurements and data analysis, LM and SC conceived the films synthesis, NM and AS performed the numerical simulations, AD and GL supported in samples preparation, RK supported in data analysis and paper writing, DG supervised the work.

## Conflicts of interest

There are no conflicts to declare.

## Acknowledgements

The authors thank the European Union under the Horizon 2020 Program, FET-Open: DNA-FAIRYLIGHTS, Grant Agreement 964995, the HORIZON-MSCA-DN-2022: DYNAMO, grant Agreement 101072818, and the Horizon Europe Program, Pathfinder Open: iSenseDNA, Grant Agreement 101046920.

## Notes and references